\documentclass[aps,prl,twocolumn,groupedaddress,superscriptaddress,floatfix,showpacs]{revtex4}

\usepackage{graphicx}
\usepackage{natbib}

\newcommand{\fig}[1]{FIG.~\ref{fig:#1}}

\newcommand{\eq}[1]{Eq.~(\ref{eq:#1})}
\newcommand{\TN}{$T_\mathrm N$}

\newcommand{\ybfo}{YBaFe$_2$O$_5$} 
\newcommand{\TV}{$T_\mathrm V$}
\newcommand{\tg}{$t_{2g}$}
\newcommand{\dxz}{$d_{xz}$}
\newcommand{\eg}{$e_g$}
\newcommand{\AFCO}{AF$_\mathrm{CO}$}
\newcommand{\AFVM}{AF$_\mathrm{VM}$}

\begin{document}

\preprint{Draft copy not for distribution -- \today}
\title{Damping of antiferromagnetic spin waves by valence fluctuations\\ in the double layer perovskite \ybfo}

\author{S. Chang} \email[]{schang@ameslab.gov}\affiliation{Ames Laboratory, Ames IA 50011}
\author{P. Karen} \affiliation{Department of Chemistry, University of Oslo, N-0315 Oslo, Norway}
\author{M.~P. Hehlen} \author{F.~R. Trouw} \affiliation{LANSCE, Los Alamos National Laboratory, Los Alamos, NM 87545}
\author{R.~J. McQueeney} \affiliation{Ames Laboratory, Ames IA 50011}\affiliation{Department of Physics and Astronomy, Iowa State University, Ames, IA 50011}

\date{\today}

\begin{abstract}
Inelastic neutron scattering experiments show that spin dynamics in the charge ordered insulating ground state of the double-layer perovskite \ybfo\ is well described in terms of \eg\ superexchange interactions.  Above the Verwey transition at \TV~=~308~K, \tg\ double exchange-type conduction within antiferromagnetic FeO$_2$--BaO--FeO$_2$ double layers proceeds by an electron hopping process that requires a spin flip of the five-fold coordinated Fe ions, costing an energy of $5 \langle J \rangle S^2 \approx 0.1$~eV. The hopping process disrupts near-neighbor spin correlations, leading to massive damping of zone-boundary spin waves.
\end{abstract}


\maketitle

The extraordinary electrical transport properties of strongly correlated transition-metal oxides are often intimately related to magnetic interactions. For example, in the doped perovskite manganites, electron spins in the itinerant \eg\ band are aligned parallel to the localized \tg\ spins due to strong Hund's rule coupling. Consequently, conduction of spin polarized \eg\ electrons results in ferromagnetic (F) nearest-neighbor interactions, known as double exchange (DE)~\cite{Zener1951}. However, antiferromagnetic (AF) superexchange (SE) via localized \tg\ electrons, cooperative Jahn--Teller distortions as well as onsite and intersite Coulomb repulsion may compete with DE leading to different insulating magnetic/orbital/charge-ordered phases~\cite{Tokura2006}. $R$BaFe$_2$O$_5$ ($R=$~Nd -- Ho and Y)~\cite{Karen2004} double layer perovskites make up another class of compounds where DE and SE interactions are in competition, resulting in either a charge-ordered insulator or valence-mixed conducting phase~\cite{Woodward2003,Karen2001}. However, in the $R$BaFe$_2$O$_5$ phases, DE occurs via the minority spin \tg\ electrons in competition with AF SE dominated by the \eg\ bands. Both SE and DE interactions may be modified in the charge/orbital-ordered ground state due to structural distortions and the fractional \dxz\ orbital occupancy at the metal site~\cite{Woodward2003}. Therefore, the study of spin excitations provides a window into the underlying charge dynamics. In this letter, we report on an inelastic neutron scattering study of the magnetic excitations in \ybfo. The charge-ordered insulating ground state below the Verwey temperature \TV\ can be well understood in terms of normal SE interactions.  However, above \TV\ a strong coupling with valence fluctuations leads to massive damping of the zone-boundary spin waves.

\begin{figure}[!tb]
	\includegraphics[width=0.35\textwidth]{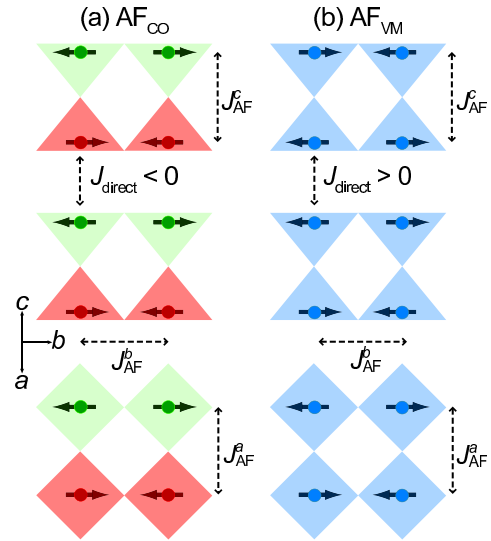}
\caption{\label{fig:ybfo_struct}(color online). Schematic representations of charge/magnetic order~\protect\cite{Woodward2003} in \ybfo. The upper and lower panels are projections onto the $bc$ and $ab$ planes, respectively. Fe ions are shown as balls and oxygen square pyramids as triangles or squares. Arrows indicate magnetic moments. (a) For $T < 308$~K (\AFCO), Fe$^{2+}$ (green online) and Fe$^{3+}$ (red online) chains run along the $b$ axis and alternate along the $a$ and $c$ directions. (b) For 308~K~$< T < 430$~K(\AFVM) the Fe ions are valence mixed (blue online).}
\end{figure}
\ybfo\ is a fractional valent material with the Fe atoms possessing a nominal valence of 2.5+. The perovskite-based crystal structure of \ybfo\ consists of FeO$_2$--BaO--FeO$_2$ double layers in which five-coordinated Fe sites form apex-shared square pyramids. The double layers are separated by an oxygen-vacant Y layer. At high temperatures, \ybfo\ is valence mixed and paramagnetic. On cooling, the compound orders antiferromagnetically at \TN~=~425~K but remains valence mixed. At \TV~=~308~K, a Verwey-type transition orders the Fe valences in real space and also modifies the AF structure. The magnetic/charge-ordering patterns~\cite{Woodward2003} for $T <$~\TV\ (\AFCO) and \TV~$< T <$~\TN\ (\AFVM) are shown in \fig{ybfo_struct}. Both \AFCO\ and \AFVM\ structures are characterized by strong AF coupling between Fe atoms within double layers due to $e_g$ SE interactions through nearly 180$^\circ$ Fe--O--Fe bonds. The main difference between the two AF structures is the sign of the direct exchange interaction between double layers (across the Y layer), which changes from F in \AFVM\ to AF in \AFCO. However, this direct exchange is much weaker than AF SE within double layers due to the large Fe--Fe interlayer distance ($\sim 3.6$~\AA~\cite{Woodward2003}). This is shown using perturbation theory to estimate the ratio of direct exchange ($J_\mathrm{direct}$) and SE ($J_\mathrm{AF}$) for a linear Fe--Fe bond~\cite{Anderson1959}
\begin{equation}\label{eq:Jd/JAF}
\frac{J_\mathrm{direct}}{J_\mathrm{AF}} \propto \frac{t_{dd}^2/U}{t_{pd}^4/\Delta^3} \approx \frac{m^2}{\hbar^4} \frac{\Delta^3}{U} \frac{\eta_{dd\sigma}^2}{\eta_{pd\sigma}^4} \frac{d_\mathrm{Fe-O}^{14}}{d_\mathrm{Fe-Fe}^{10}} = 1\%,
\end{equation}
where $t_{pd}$ and $t_{dd}$ are the Fe--O and Fe--Fe overlap integrals, $U$ is the onsite Coulomb repulsion, and $\Delta$ is the charge transfer energy. These parameters are estimated as $U = 8$~eV and $\Delta=3$~eV~\cite{mizokawa1998}. 
Overlap integrals are estimated using Harrison's method~\cite{Harrison} with $d_\mathrm{Fe-Fe} = 3.6$~\AA, $d_\mathrm{Fe-O}=2.0$~\AA~\cite{Woodward2003}, and the coefficients of the radial overlap integrals are $\eta_{dd\sigma}=16.2$ and $\eta_{dp\sigma}=2.95$~\cite{Harrison}.

To study the spin waves, we performed time-of-flight (TOF) powder inelastic neutron scattering measurements on approximately 50~g of \ybfo\ using the Pharos spectrometer at the Lujan Center, Los Alamos National Laboratory. The master sample of \ybfo\ was synthesized in several batches from citrate precursors as reported in Ref.~\cite{Woodward2003}. The combined batches were equilibrated at 400$^\circ$~C for 33 days in a sealed silica ampoule together with 0.03~g Zr foil as a getter, placed 12~cm from the sample and locally heated to 700$^\circ$~C. This was followed by a cool down at a rate of 0.1$^\circ$~C/min. The cerimetrically determined oxygen content was 5.005(1) per formula at the top of this batch and 5.003(1) at the bottom. The sample quality was verified by X-ray powder diffraction, differential scanning calorimetry, and magnetization measurements, which were in good agreement with previous results~\cite{Woodward2003}.  The powder sample was loaded in a flat aluminum can and mounted on the cold head of a closed-cycle He cryostat. For temperatures greater than 325~K, a displex with a high temperature stage was used.  The sample was oriented at 45$^\circ$ to the incident neutron beam and inelastic neutron spectra were measured at various temperatures between 6~K and 450~K, with an incident energy $E_i=120$~meV. The TOF data were corrected for energy dependent detector efficiencies and instrumental background and reduced into energy transfer ($E$) and scattering angle ($2\theta$) histograms. The resulting spectra contain both magnetic and phonon contributions. However, since the magnetic intensity disappears at high angles due to the magnetic form factor, it is possible to segregate the magnetic and phonon scattering by summing the data over different angle ranges. For the present experiment, we chose angle ranges $2\theta=1-35^\circ$ and $55-95^\circ$ for low (magnetic) and high (phonon) angle data, respectively. The phonon contribution to the scattered intensity in the low angle data was removed by subtracting the appropriately scaled high angle data.

\begin{figure}[!tb]
  \includegraphics[width=0.35\textwidth]{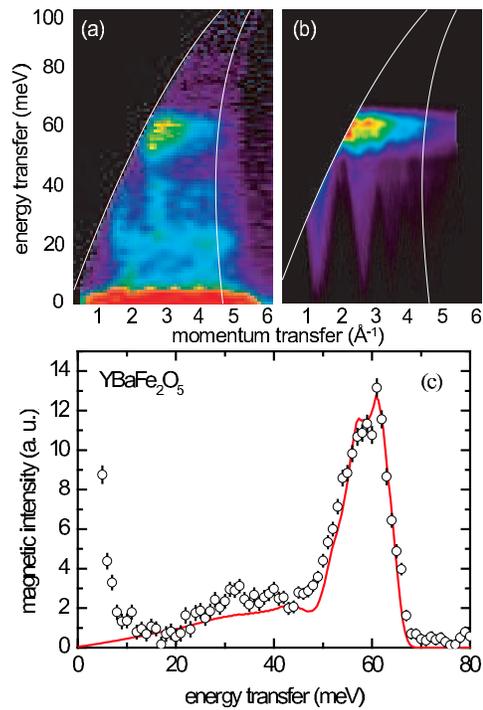}\\
\caption{\label{fig:sqeM_base}(color online). (a) Inelastic neutron scattering intensity, $S(Q,E)$, for \ybfo\ at $T=6$~K. White lines are loci of constant angles in ($Q,E$)-space that denote limits of the angle summation, $2\theta = 1^\circ - 35^\circ$. (b) Calculation of $S(Q,E)$ for \ybfo\ at $T = 6$~K using a Heisenberg model with $J_{33}  = 5.9$~meV, $J_{22} = 3.4$~meV and $J_{23} = 6.0$~meV. (c) Comparison of angle-summed magnetic scattering data (open symbols) with Heisenberg model calculation (line).}
\end{figure}
The total scattered intensity for \ybfo\ at 6~K in the charge ordered \AFCO\ phase is shown as a function of $Q$ (the magnitude of momentum transfer vector $\mathbf Q$) and $E$ in \fig{sqeM_base}(a). The main feature in the spectrum in the angle range $1^\circ - 35^\circ$ is a prominent band of scattering between $50 - 65$~meV arising from zone-boundary spin waves.  The band appears as two (or more) overlapping peaks in the phonon subtracted magnetic intensity shown in \fig{sqeM_base}(c).  Additional weak features below 40~meV are due to imperfect subtraction of phonon bands near 20~meV and 40~meV, as well as dispersive features from both spin waves and phonons appearing as vertical streaks in \fig{sqeM_base}(a).

\begin{figure}[!tb]
  \includegraphics[width=0.35\textwidth]{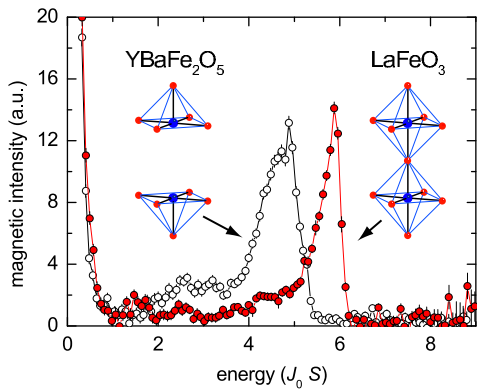}
\caption{\label{fig:ybfo_lfo}(color online). Magnetic scattering intensity for \ybfo\ at 6~K and LaFeO$_3$~\protect\cite{Mcqueeney2006b} at 10~K. Insets show idealized oxygen square pyramids(octahedra) in \ybfo(LaFeO$_3$) illustrating the broken magnetic bond in \ybfo\ in comparison to LaFeO$_3$.}
\end{figure}
Measurements of polycrystalline-averaged spin excitation spectra of long-range magnetically ordered material result in neutron scattering intensities related to the spin-wave density-of-states (SWDOS).  At low temperatures, the SWDOS of insulating and charge-ordered \ybfo\ may be understood by considering only SE interactions~\cite{Anderson1959} 
within double layers and is comparable to the G-type AF perovskite LaFeO$_3$ (see \fig{ybfo_lfo}).  LaFeO$_3$ may be considered isotropic with a single nearest-neighbor exchange constant $J_0 \approx 5$~meV and spin $S = 5/2$~\cite{Mcqueeney2006b}. Accordingly, the SWDOS of LaFeO$_3$ consists of a single sharp peak indicating a zone-boundary spin-wave energy of $6 J_0 S$ ($\sim$75~meV) due to the octahedral coordination of the Fe sites. The average zone-boundary spin-wave energy for \ybfo\ ($\sim$60~meV) is very close to $5 J_0 S$. Therefore, the SWDOS in \ybfo\ is comparable to LaFeO$_3$, after accounting for the one broken AF exchange bond (five nearest neighbors) present in the square-pyramidal coordination.

While the comparison of the \AFCO\ spectrum to that of LaFeO$_3$ shows that the energy scale of the SWDOS in \ybfo\ is set by the average exchange $\langle J \rangle \approx J_0$, details such as the appearance of split peaks (see \fig{sqeM_base}) can only be explained by taking into account the variations in the exchange between different Fe--Fe pairs. As illustrated in \fig{ybfo_struct}(a), charge order in the \AFCO\ phase consists of Fe$^{2+}$ and Fe$^{3+}$ chains along the $b$ axis, resulting in four unique nearest-neighbor SE paths, given by $J^b_{33}$, $J^b_{22}$, $J^a_{23}$, and $J^c_{23}$, where the subscripts indicate the valences of the Fe--Fe pairs and the superscripts denote the direction of the magnetic bonds. It is reasonable to assume that the SE integral can be transferred from LaFeO$_3$ to the Fe$^{3+}$--Fe$^{3+}$ pair in \ybfo\ using the approximate relationship $J_{33}^b \propto \frac{1}{S^2} t^4 \cos^2 \theta$~\cite{Mishra1998}, 
where $t$ is the Fe$^{3+}$--O transfer integral (for 180$^\circ$ bond angles) which depends sensitively on the Fe--O distance $d$ as $t \propto d^{-7/2}$~\cite{Harrison} and $\theta$ is the Fe--O--Fe bond angle. We have used the resulting estimate for $J_{33}$ (5.9~meV) as a starting point for studies of model calculations using the Heisenberg Hamiltonian
\begin{equation}\label{eq:hamiltonian}
H = \sum_{\langle i,j \rangle} J_{ij} \mathbf S_i \cdot \mathbf S_j,
\end{equation}
where $J_{ij}$ is the exchange energy between spins $\mathbf S_i$ and $\mathbf S_j$ and $\langle i,j \rangle$ indicates that the sum is only over nearest neighbors. The Heisenberg Hamiltonian given in \eq{hamiltonian} was used within linear spin-wave theory to obtain spin-wave energies and eigenvectors from which neutron intensities due to coherent scattering $S(\mathbf Q, E)$ were calculated. For a more detailed discussion of model spin-wave calculations, see Ref.~\cite{McQueeney2006a}. Polycrystalline averaging of $S(\mathbf Q, E)$ was performed by Monte-Carlo integration over $10000$ $\mathbf Q$-vectors on each constant-$Q$ sphere. The model calculations show that the two peak energies in \ybfo\ (at 58 and 62 meV) depend on 
\begin{eqnarray}\label{splitting}
E_2 & \approx & 2J_{22}^b S_2 + \left(2J_{23}^a + J_{23}^c\right) S_3 \nonumber\\
E_3 & \approx & 2J_{33}^b S_3 + \left(2J_{23}^a + J_{23}^c\right) S_2,
\end{eqnarray}
where $S_i$ are $5/2$ and 2 for Fe$^{3+}$ or Fe$^{2+}$, respectively. These two energies are understood as the local excitation energy of the Fe$^{3+}$ (Fe$^{2+}$) spin inside the respective square pyramid.  The combination $J_{23} = \left(2J_{23}^a + J_{23}^c\right)/3$ shifts the center-of-mass of the entire spin-wave band and $J_{22}^b$ controls the splitting of the two peaks.  Best agreement with both the splitting and the intensity ratio of the two peaks results in $J_{33}^b \approx 5.9$~meV, $J_{22}^b \approx 3.4$~meV, and $J_{23} \approx 6.0$~meV for the charge-ordered state \AFCO. The calculated powder-averaged magnetic neutron intensities $S(Q,E)$ and $S(E)$, shown in \fig{sqeM_base}(b) and (c), respectively are in excellent agreement with the measured data.

\begin{figure}[!tb]
  \includegraphics[width=0.35\textwidth]{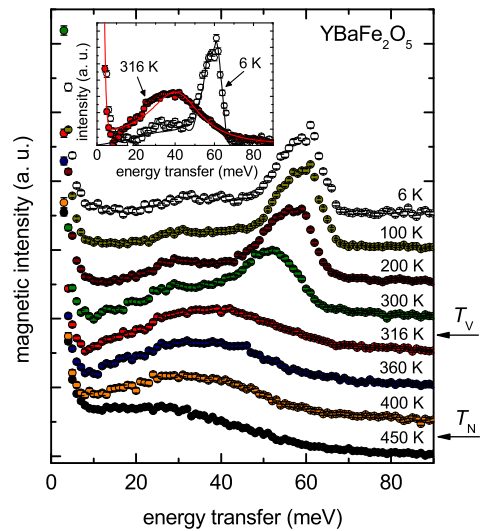}
\caption{\label{fig:se_waterfall}(color online). Temperature dependence of the magnetic scattering intensity for \ybfo\ between 6~K and 450~K.  Intensities at each temperature have been offset by a uniform amount for clarity. Spectra at 6~K (\AFCO) and 316~K (\AFVM) are compared in the inset. The lines are results of a fit to a damped harmonic oscillator model(316~K data) and a Heisenberg model calculation (6~K data).}
\end{figure}
\fig{se_waterfall} shows the magnetic intensity as a function of temperature from 6~K to 450~K.  Below \TV~=~308~K, the main spin-wave band is at 60~meV, as discussed above, although there is some broadening and softening in energy at 300~K, just below \TV.  At 316~K, just above \TV\, in the \AFVM\ phase, a dramatic shift in the energy of the spin-wave peak down to 40~meV is accompanied by a huge broadening of the spin-wave band. Above \TN~=~450~K, paramagnetic scattering is observed. The spectra in the \AFCO\ and \AFVM\ phases are compared in the inset of \fig{se_waterfall}. The large decrease of zone-boundary spin-wave energy and the massive broadening in the valence mixed \AFVM\ phase do not arise from normal SE interactions and can not be reproduced by a simple Heisenberg model. The broadening may arise from a distribution of exchange along different paths and/or from the presence of overdamped modes. In the \AFVM\ phase, the valence-mixed Fe ions are electronically equivalent and occupy a single crystallographic site with Fe--O--Fe bond distances nearly identical in all directions.  Thus, a decrease in the spin-wave bandwidth is expected in the \AFVM\ phase in comparison to \AFCO.

The flipping of the sign of the direct exchange from AF (\AFCO) to F (\AFVM) at \TV\ might appear to have a significant effect on the SWDOS. However, introduction of the direct exchange $J_\mathrm{direct}$ to model spin-wave calculations leads to a splitting of the SWDOS proportional to $J_\mathrm{direct}$ rather than a decrease in energy of the entire spectrum.  This confirms that $J_\mathrm{direct}$ does not have a significant effect on the SWDOS, as was already expected from \eq{Jd/JAF}. Therefore, the broadening and decrease in energy of the zone-boundary spin waves suggest damping arising from a strong coupling of spin excitations with valence-mixed charge carriers.

The large changes in the spin-wave spectrum at \TV\ are explained by the appearance of new ferromagnetic interactions within the double layers.  Valence-mixed \tg\ bands give rise to ferromagnetic DE that competes with \eg\ AF SE. In doped perovskite manganites, strong Hund's rule coupling of the itinerant \eg\ electron to the localized \tg\ spin overwhelms the AF SE, leading to ferromagnetism via DE~\cite{Brink1999}. In other words, metallic ferromagnetism in the manganites is established in the limit $J_\mathrm{DE} > J_\mathrm{SE}$, where $J_\mathrm{DE}$ and $J_\mathrm{SE}$ represent DE and SE, respectively within a Heisenberg model. To estimate the strength of the DE within double layers, the spin-wave spectrum at 316~K in \AFVM\ was fit using a damped harmonic oscillator model
\begin{equation}\label{DHO}
S(E) = \frac{1}{1 - e^{-E/k_BT}}\frac{S_0 E \Gamma}{(E^2-E_0^2)^2 + (E\Gamma)^2},
\end{equation}
where $(1 - e^{-E/k_BT})^{-1}$ is the Bose factor, $\Gamma = 33(1)$~meV is the width and $E_0=45.8(4)$~meV is the position. The fit is shown as a red line in the inset of \fig{se_waterfall}. We estimate the average exchange energy from $E_0 = 5\langle J \rangle S$, resulting in $\langle J \rangle \approx 4.07$~meV; a 23\% decrease from \AFCO. Given that the average exchange energy in the valence-mixed \AFVM\ phase of \ybfo\ is $\langle J \rangle = J_\mathrm{SE}+J_\mathrm{DE} \approx 4.07$~meV and $J_\mathrm{SE} \approx 5.3$~meV in the charge-ordered \AFCO\ phase, we estimate that $J_\mathrm{DE} \approx -1.2$~meV and $|J_\mathrm{DE}/J_\mathrm{SE}| \approx 0.23$. In contrast to the manganites, conduction in \ybfo\ proceeds via the minority spin \tg\ electron, whose spin must be opposite that of the local spin on the half-filled Fe$^{3+}$ ion by the Pauli exclusion principle. Therefore, spin-polarized conduction within a double layer must overcome the AF structure established by the stronger, \eg\ dominated SE.  This can occur by flipping the spin on one Fe ion, momentarily creating an arrangement of five neighboring Fe spins that are ferromagnetically aligned the central spin. It is then possible for real hopping of the \tg\ electron to occur within a double layer through a DE-type process. Flipping the spin at one Fe site costs an energy of $5\langle J \rangle S^2 \approx 0.1$~eV. This scenario is consistent with conductivity measurements in GdBaFe$_2$O$_5$~\cite{Linden2006} indicating that transport is an activated process in the valence mixed phase, with an activation energy of about 0.1~eV. Such a hopping process will strongly disrupt near-neighbor spin correlations, leading to massive damping of zone-boundary spin waves. From the perspective of the zone-boundary spin waves, DE implies that deviations of the relative angle between neighboring spins makes electron hopping easier~\cite{Tokura2006}, causing the decay of the zone-boundary spin wave into a charge and multi-magnon excitations. Given such a picture, a significant change in the activation energy should be observed in resistivity in applied magnetic fields or at \TN.

\begin{acknowledgments}
We thank J.~Q. Yan, A. Kreyssig and P.~M. Woodward for their input. Ames Laboratory is supported by the U.S. Department of Energy Office of Science under Contract No.~DE-AC02-07CH11358. This work has benefited from the use of the Los Alamos Neutron Science Center at Los Alamos National Laboratory. LANSCE in funded by the U.S. Department of energy under Contract No.~W-7405-ENG-36.
\end{acknowledgments}



\begin{thebibliography}{16}
\expandafter\ifx\csname natexlab\endcsname\relax\def\natexlab#1{#1}\fi
\expandafter\ifx\csname bibnamefont\endcsname\relax
  \def\bibnamefont#1{#1}\fi
\expandafter\ifx\csname bibfnamefont\endcsname\relax
  \def\bibfnamefont#1{#1}\fi
\expandafter\ifx\csname citenamefont\endcsname\relax
  \def\citenamefont#1{#1}\fi
\expandafter\ifx\csname url\endcsname\relax
  \def\url#1{\texttt{#1}}\fi
\expandafter\ifx\csname urlprefix\endcsname\relax\def\urlprefix{URL }\fi
\providecommand{\bibinfo}[2]{#2}
\providecommand{\eprint}[2][]{\url{#2}}

\bibitem[{\citenamefont{Zener}(1951)}]{Zener1951}
\bibinfo{author}{\bibfnamefont{C.}~\bibnamefont{Zener}},
  \bibinfo{journal}{Phys. Rev.} \textbf{\bibinfo{volume}{82}},
  \bibinfo{pages}{403} (\bibinfo{year}{1951}).

\bibitem[{\citenamefont{Tokura}(2006)}]{Tokura2006}
\bibinfo{author}{\bibfnamefont{Y.}~\bibnamefont{Tokura}},
  \bibinfo{journal}{Rep. Prog. Phys.} \textbf{\bibinfo{volume}{69}},
  \bibinfo{pages}{797} (\bibinfo{year}{2006}).

\bibitem[{\citenamefont{Karen}(2004)}]{Karen2004}
\bibinfo{author}{\bibfnamefont{P.}~\bibnamefont{Karen}}, \bibinfo{journal}{J.
  Solid State Chem.} \textbf{\bibinfo{volume}{177}}, \bibinfo{pages}{281}
  (\bibinfo{year}{2004}).

\bibitem[{\citenamefont{Woodward and Karen}(2003)}]{Woodward2003}
\bibinfo{author}{\bibfnamefont{P.~M.} \bibnamefont{Woodward}} \bibnamefont{and}
  \bibinfo{author}{\bibfnamefont{P.}~\bibnamefont{Karen}},
  \bibinfo{journal}{Inorg. Chem.} \textbf{\bibinfo{volume}{42}},
  \bibinfo{pages}{1121} (\bibinfo{year}{2003}).

\bibitem[{\citenamefont{Karen et~al.}(2001)}]{Karen2001}
\bibinfo{author}{\bibfnamefont{P.}~\bibnamefont{Karen}} \bibnamefont{et~al.},
  \bibinfo{journal}{Phys. Rev. B} \textbf{\bibinfo{volume}{64}},
  \bibinfo{pages}{214405} (\bibinfo{year}{2001}).

\bibitem[{\citenamefont{Anderson}(1959)}]{Anderson1959}
\bibinfo{author}{\bibfnamefont{P.~W.} \bibnamefont{Anderson}},
  \bibinfo{journal}{Phys. Rev.} \textbf{\bibinfo{volume}{115}},
  \bibinfo{pages}{2} (\bibinfo{year}{1959});
  \bibinfo{author}{\bibfnamefont{J.~B.} \bibnamefont{Goodenough}},
  \bibinfo{journal}{\emph{ibid}} \textbf{\bibinfo{volume}{171}},
  \bibinfo{pages}{466} (\bibinfo{year}{1968}).

\bibitem[{\citenamefont{Mizokawa and Fujimori}(1998)}]{mizokawa1998}
\bibinfo{author}{\bibfnamefont{T.}~\bibnamefont{Mizokawa}} \bibnamefont{and}
  \bibinfo{author}{\bibfnamefont{A.}~\bibnamefont{Fujimori}},
  \bibinfo{journal}{Phys. Rev. Lett.} \textbf{\bibinfo{volume}{80}},
  \bibinfo{pages}{1320} (\bibinfo{year}{1998}); 
  \bibinfo{author}{\bibfnamefont{A.~E.} \bibnamefont{Bocquet}}
  \bibnamefont{et~al.}, \bibinfo{journal}{Phys. Rev. B}
  \textbf{\bibinfo{volume}{45}}, \bibinfo{pages}{1561} (\bibinfo{year}{1992}).

\bibitem[{\citenamefont{Harrison}(1989)}]{Harrison}
\bibinfo{author}{\bibfnamefont{W.~A.} \bibnamefont{Harrison}},
  \emph{\bibinfo{title}{Electronic Structure and the Properties of Solids}}
  (\bibinfo{publisher}{Dover}, \bibinfo{address}{New York},
  \bibinfo{year}{1989}).

\bibitem[{\citenamefont{McQueeney et~al.}(2006{\natexlab{a}})}]{Mcqueeney2006b}
\bibinfo{author}{\bibfnamefont{R.~J.} \bibnamefont{McQueeney}}
  \bibnamefont{et~al.}, \eprint{cond-mat} (\bibinfo{note}{0607570, 2006}).

\bibitem[{\citenamefont{Mishra and Satpathy}(1998)}]{Mishra1998}
\bibinfo{author}{\bibfnamefont{S.~K.} \bibnamefont{Mishra}} \bibnamefont{and}
  \bibinfo{author}{\bibfnamefont{S.}~\bibnamefont{Satpathy}},
  \bibinfo{journal}{Phys. Rev. B} \textbf{\bibinfo{volume}{58}},
  \bibinfo{pages}{7585} (\bibinfo{year}{1998});
  \bibinfo{author}{\bibfnamefont{H.}~\bibnamefont{Meskine}},
  \bibinfo{author}{\bibfnamefont{H.}~\bibnamefont{K\"onig}}, \bibnamefont{and}
  \bibinfo{author}{\bibfnamefont{S.}~\bibnamefont{Satpathy}},
  \bibinfo{journal}{\emph{ibid}} \textbf{\bibinfo{volume}{64}},
  \bibinfo{pages}{094433} (\bibinfo{year}{2001}).

\bibitem[{\citenamefont{McQueeney et~al.}(2006{\natexlab{b}})}]{McQueeney2006a}
\bibinfo{author}{\bibfnamefont{R.~J.} \bibnamefont{McQueeney}}
  \bibnamefont{et~al.}, \bibinfo{journal}{Phys. Rev. B}
  \textbf{\bibinfo{volume}{73}}, \bibinfo{pages}{174409}
  (\bibinfo{year}{2006}{\natexlab{b}}).

\bibitem[{\citenamefont{v.~d. Brink and Khomskii}(1999)}]{Brink1999}
\bibinfo{author}{\bibfnamefont{J.}~\bibnamefont{v.~d. Brink}} \bibnamefont{and}
  \bibinfo{author}{\bibfnamefont{D.}~\bibnamefont{Khomskii}},
  \bibinfo{journal}{Phys. Rev. Lett.} \textbf{\bibinfo{volume}{82}},
  \bibinfo{pages}{1016} (\bibinfo{year}{1999}).

\bibitem[{\citenamefont{Lind\'en et~al.}(2006)}]{Linden2006}
\bibinfo{author}{\bibfnamefont{J.}~\bibnamefont{Lind\'en}}
  \bibnamefont{et~al.}, \bibinfo{journal}{Phys. Rev. B}
  \textbf{\bibinfo{volume}{73}}, \bibinfo{pages}{064415}
  (\bibinfo{year}{2006}).

\end{thebibliography}
\end{document}